\documentclass[10pt,conference]{IEEEtran}

\usepackage{flushend}

\usepackage[utf8]{inputenc}

\usepackage[
    style=ieee,
    sorting=nyt,
]{biblatex}

\bibliography{agse,dpp,general}

\DeclareSourcemap{\maps[datatype=bibtex]{
\map{
    \step[fieldsource=booktitle, match={Conference}, replace={Conf.}]
    \step[fieldsource=booktitle, match={Symposium}, replace={Symp.}]
    \step[fieldsource=booktitle, match={Proceedings of the}, replace={Proc.}]
    \step[fieldsource=booktitle, match={Proceedings}, replace={Proc.}]
    \step[fieldsource=booktitle, match={International}, replace={Intl.}]
    \step[fieldsource=journal, match={International Journal}, replace={Intl. J.}]        
}%
}}

\usepackage{hyperref}

\newenvironment{excerpt}%
{%
    \list{}{\leftmargin=2em\rightmargin=1.5em}%
    \small%
    \setlength{\parindent}{-1.6em}%
    \item[]%
    \hspace{-2em}%
}%
{\endlist}

\usepackage{xcolor}

\definecolor{devOneClr}{RGB}{143, 27,  129}
\definecolor{devTwoClr}{RGB}{ 29, 101, 129}

\newcommand{\qa}[2]{%
    #1: \textcolor{devOneClr}{``#2''}%
}
\newcommand{\qb}[2]{%
    #1: \textcolor{devTwoClr}{``#2''}%
}

\usepackage{xspace}
\newcommand{\cOne}{\textsl{C1}\xspace}
\newcommand{\cTwo}{\textsl{C2}\xspace}
\newcommand{\dThree}{\textsl{D3}\xspace}
\newcommand{\dFour}{\textsl{D4}\xspace}
\newcommand{\jOne}{\textsl{J1}\xspace}
\newcommand{\jTwo}{\textsl{J2}\xspace}
\newcommand{\pOne}{\textsl{P1}\xspace}
\newcommand{\pThree}{\textsl{P3}\xspace}

\begin{document}

\title{Two Elements of Pair Programming Skill}

\author{\IEEEauthorblockN{Franz Zieris}
\IEEEauthorblockA{\textit{Institut für Informatik} \\
\textit{Freie Universität Berlin}\\
Berlin, Germany \\
zieris@inf.fu-berlin.de}
\and
\IEEEauthorblockN{Lutz Prechelt}
\IEEEauthorblockA{\textit{Institut für Informatik} \\
\textit{Freie Universität Berlin}\\
Berlin, Germany \\
prechelt@inf.fu-berlin.de}}

\maketitle

\begin{abstract}
    \textit{Background:}
    Pair programming (PP) can have many benefits in industry.
    Researchers and practitioners recognize that successful and productive PP
    involves some \emph{skill} that might take time to learn and improve.

    \textit{Question:}
    What are the elements of pair programming skill?

    \textit{Method:}
    We perform qualitative analyses of industrial pair programming
    sessions following the Grounded Theory Methodology.
    We look for patterns of problematic behavior to conceptualize key elements
    of what `good' and `bad' pairs do differently.

    \textit{Results:}
    Here, we report two elements of pair programming skill:
    Good pairs (1)~manage to maintain their \emph{Togetherness} and
    (2) keep an eye on their session's \emph{Expediency}.
    We identify three problematic behavioral patterns that affect one or both
    of these elements:
    \emph{Getting Lost in the Weeds}, \emph{Losing the Partner}, and
    \emph{Drowning the Partner}.

    \textit{Conclusion:}
    Pair programming skill is separate from general software development skill.
    Years of PP experience are neither a prerequisite nor sufficient
    for successful pair programming.
\end{abstract}

\section{Introduction}

\emph{Pair programming} (PP) is the practice of two software developers working
closely together on one machine.
Kent Beck characterizes it as \emph{``a dialog between two people trying to
[...] program (and analyze and design and test)''}
which \emph{``is a subtle skill, one that you can spend the rest of your life
getting good at''} \cite[p.~100]{Beck99}.
Beck sees many benefits in this practice, such as higher code quality
in less time \cite[pp.~66--67]{Beck99}.
He does not, however, elaborate on the aspects of the ``skill'' underlying
these benefits; he merely alludes to the importance of
\emph{communication} and \emph{coordination} \cite[p.~141]{Beck99}.

Much of pair programming research appears to be built on the assumption
that PP does \emph{not} involve any particular skill beyond general software
development experience:
A large experiment by Arisholm et al.\ \cite{AriGalDyb07}, for instance,
was set to determine PP's effect on code quality and effort for junior,
intermediate, and expert developers, but
93 of its 98 subject pairs had \emph{no prior pairing experience at all}.
A meta-analysis of PP effectiveness \cite{HanDybAri09} later found only weak effects and
high between-study variance, which indicates a number of not-understood
(and hence uncontrolled) moderating factors including the
\emph{``amount of training in pair programming''} \cite[Sec.~4]{HanDybAri09},
which the researchers expect to have a positive effect on pair performance.

To understand the differences between skillful and problematic pair
programming, we perform qualitative analyses of industrial PP sessions.
Here, we report two elements of \emph{PP skill} we identified.
We discuss related work in Section~\ref{sec:related-work},
characterize our data and explain our research method in
Section~\ref{sec:data-method},
sketch our findings in Section~\ref{sec:results},
and provide a discussion and an outlook in Sections~\ref{sec:discussion}
and \ref{sec:conclusion-further-work}.

\section{Related Work}
\label{sec:related-work}

\subsection{On the Existence of Pair Programming Skill}

In the context of her PP experiments with students, Williams \cite{Williams00}
describes the process of \emph{pair jelling} as the transition of two
individuals from
\emph{``considering themselves as a two-pro\-gram\-mer team [to] considering
themselves as one coherent, intelligent organism working with one mind''}
\cite[p.~53]{Williams00}.
Such \emph{jelling} appears to have two aspects:
Individuals get accustomed to not working
alone \cite[p.~22]{WilKesCun00},
and a specific pair gets to know each other in order for some
\emph{``bonding''} to happen \cite[p.~101]{Williams00}.
Williams does not explicate what this phase entails or how long it takes,
but she claims the student pairs were jelled after their first
experiment assignment \cite[pp.~63--64]{Williams00}.

From an industry context, practitioner Belshee reports that
\emph{``[i]t often takes days for a given pair to be comfortable with each
other''},
which he describes as a precondition for a pair being able to reach a state
of highly productive \emph{``Pair Flow''} in which both members have a shared
understanding of their task \cite[Sec.~1.2]{Belshee05}.
Belshee does not, however, mention \emph{what} or \emph{how long} it takes to get
good at pair programming.

Bryant et al.\ \cite{Bryant04,BryRomBou08} analyze the abstraction level of
utterances in the dialog of pair programmers in industry and note that there
are differences between the pair member sitting at the keyboard and her
partner---%
but \emph{only} for pairs with less than six months of PP experience:
The frequencies and types of utterances of \emph{``expert pair programmers''}
do not depend on who controls the keyboard \cite[Sec.~5.2]{Bryant04}.
Bryant et al.\ conclude that in experienced pairs, both partners
maintain a \emph{``clear mental model of [the pair's] current state''}
\cite[Sec.~6.3]{BryRomBou08}.

\subsection{On the Elements of Pair Programming Skill}

While there is some awareness in the literature that two developers are not
suddenly more productive just because they are placed next to each other,
there has been only little research into the elements that make pair
programming work.
In their ethnographic study, 
Chong \& Hurlbutt \cite{ChoHur07} report that two partners with a 
similar level of task-relevant expertise can become \emph{``tightly coupled''} 
such that they do not even need complete sentences for effective communication.
However, with a large-enough expertise difference between them,
the more knowledgable partner will be dominant.

For the aspect of transferring knowledge during pair programming,
both Plonka et al.\ \cite{PloShaLin15} and
Zieris \& Prechelt \cite{ZiePre14-ppknowtrans,ZiePre16-ppknowtrans2}
summarize a number of `expertly' pair programming behaviors, including:
Making one's thinking process visible
\cite{PloShaLin15,ZiePre16-ppknowtrans2},
letting the learner work something out on her own \cite{PloShaLin15},
letting the explainer finish her thoughts \cite{ZiePre14-ppknowtrans},
and making sure to deal with complex topics completely
\cite{ZiePre14-ppknowtrans}.

\section{Research Method}
\label{sec:data-method}

\subsection{Research Goal and Data Collection}

The overall goal of our research is to understand how `good' and
`bad' pair programming sessions differ.
Ultimately, we want to provide actionable advice for practitioners.
Here, we want to understand the \emph{elements} of the \emph{skill} which pair
programmers exhibit in successful sessions and \emph{how} sessions suffer from
a lack thereof.

The industrial data used by Bryant et al.\ \cite{Bryant04,BryRomBou08} is
limited to audio recordings, which makes it difficult to
understand what the developers are referring to:
For one out of every eight utterances, the researchers could not reconstruct
what the pairs referred to \cite[Sec.~5.1]{BryRomBou08}.
We therefore analyze industrial PP sessions comprising \emph{audio},
\emph{webcam}, and \emph{screencast} from the \emph{PP-ind} repository
\cite{ZiePre20-ppind-tr,PPind-1.0},
which contains a variety of over 60 everyday PP sessions from 13 companies
along with pre- and post-session questionnaires filled out by the developers.
Sessions from the repository have IDs like `\textsl{CA2}'
(session 2, from the first team A, at the third company C); developers are numbered similarly,
e.g. `\cTwo'.

\subsection{Qualitative Research Approach}

We follow Strauss' \& Corbin's Grounded Theory Methodology \cite{StrCor90}.
In particular, we perform \emph{theoretical sampling} \cite[Ch.~11]{StrCor90}
by choosing sessions from the repository with pair members who have been
pair-programing regularly for years and those that are new to the practice,
as well as involving experienced software developers and novices
(see Table~\ref{tab:data}).
Below, we report our findings mostly from \emph{open coding}
\cite[Ch.~5]{StrCor90}, where relevant phenomena in the data are identified,
analyzed, and characterized through concepts,
and some findings from \emph{axial coding} \cite[Ch.~7]{StrCor90}, which
investigates when and how these phenomena occur and
how the pair deals with them.
We did not yet perform \emph{selective coding} \cite[Ch.~8]{StrCor90} to
formulate a theory and also did not yet reach \emph{theoretical saturation}
\cite[p.~188]{StrCor90}.

\begin{table}
    \caption{PP sessions analyzed from the \emph{PP-ind} repository
        \cite{ZiePre20-ppind-tr}}
    \label{tab:data}
    \begin{tabular}{
        @{}c@{\hspace{4pt}}c@{\hspace{2pt}}|
        @{\hspace{2pt}}c@{\hspace{4pt}}c@{\hspace{4pt}}c@{\hspace{2pt}}|
        @{\hspace{2pt}}c@{\hspace{4pt}}c@{\hspace{4pt}}c@{\hspace{2pt}}|
        @{\hspace{2pt}}p{2.9cm}@{}
    }
        \textbf{Session} & \textbf{Length} &
        \multicolumn{3}{@{\hspace{2pt}}c@{\hspace{2pt}}|@{\hspace{2pt}}}{\textbf{Member \#1}} &
        \multicolumn{3}{@{\hspace{2pt}}c@{\hspace{2pt}}|@{\hspace{2pt}}}{\textbf{Member \#2}} &
        \multicolumn{1}{c}{\textbf{Session content}} \\
        ID & h:mm & ID & Dev. & PP & ID & Dev. & PP
        \\
        \hline
        \textsl{CA1} & 1:18 & \cOne & 4y & 2y & \cTwo & 9y & 6y &
        Implementation of a new GUI form (\cOne already started) \\\hline
        \textsl{DA2} & 2:24 & \dThree & 3m & 3m & \dFour & 1y & 0 &
        Feature implementation pivoted to refactoring
        (\dFour's 1\textsuperscript{st} PP ever) \\\hline
        \textsl{JA1} & 1:07 & \jOne & 6y & $>$6m & \jTwo & 2.5y &
        $>$6m & Walkthrough of \jTwo's code, discussion of possible
        refactorings \\\hline
        \textsl{PA3} & 1:31 & \pOne & 5y & 2y & \pThree & 5y & 2y
        & Implement new API endpoint incl.\ testing (\pThree already
        started) \\
    \end{tabular}
\end{table}

\subsection{Our Notions of `Good' and `Bad'}
\label{sec:good-and-bad}

We only assess \emph{exhibited} PP skill, not the developers' potential
and not behavior changes over a longer time.
This is also a purely qualitative study.
We use a deficit-oriented perspective, i.e., we analyze
episodes of pairs running into `trouble', e.g.,
(a)~one or both pair members getting frustrated because they do not understand
what their partner says or does, or
(b)~the pair doing things that help neither with their actual task
nor with some overarching goal such as getting familiar with the code base.

\section{Results}
\label{sec:results}

We first describe two elements of pair programming skill
(\emph{Togetherness} and \emph{Expediency})
and then characterize
three patterns of problematic behavior in pair programming sessions:
(\emph{Getting Lost in the Weeds},
\emph{Losing the Partner}, and
\emph{Drowning the Partner}).
We illustrate each pattern with excerpts from actual sessions.

Section~\ref{sec:positive} then gives examples of the alternate (high-skill) case,
where pairs manage to avert these problems.

\subsection{Two Elements of Pair Programming Skill}
\label{sec:two-elements}

\textbf{Togetherness.}
Good pairs manage to stay together, that is,
to establish and maintain a shared mental model throughout
their session.
They detect and address relevant discrepancies in each other's understanding of
their task, work state, software system, and software development in general.

\textbf{Expediency.}
Good pairs balance short-term goals, such as identifying a defect or
implementing a new feature, and long-term goals,
e.g., addressing any member's knowledge gaps.

\subsection{Anti-Pattern: Getting Lost in the Weeds}
\label{sec:pat-weeds}

Two developers may come up with more ideas on what to look up and how to
proceed than a single developer.
But pairs risk \emph{Getting Lost in the Weeds} when they
jump on too many of them with too little consideration.
Such pairs may manage to \emph{Stay Together} in that they both think about all
these new ideas together, but they risk thinking too much about
irrelevant details, losing track of what is important, and thus reduce their
\emph{Expediency}.

\textbf{Example 1: Session \textsl{DA2} (09:00--19:00).}
It is developer \dFour's first week at the company, and he and \dThree
are tasked with implementing a new feature.
\dThree wants to explain the target state by showing a similar, already
existing function.
While \dThree scrolls through the source code, \dFour repeatedly
interrupts him with questions unrelated to their task, and \dThree
always tries his best to provide all the information he can
(highly compressed excerpt follows):

\begin{excerpt}
    \qa{\dThree}{In principle, there should be a toolbar up here [...]
        I’ll show you how it looked in the old calendar.
        [starts navigating in the source code]}

    \qb{\dFour}{[reading from screen] What are these navigation things for?
        Are these \emph{Actions}? Where are they displayed?}

    \qa{\dThree}{[stops navigating] There is a---What’s it called again?
        [starts searching through package tree ...]}

    \qb{\dFour}{[later: reading from screen, chuckling] LicenseKey?}

    \qa{\dThree}{[stops searching] You'll see that more often around here
        [...] let's see where it's used [starts fulltext search ...]}
\end{excerpt}

Since \dFour is new at the company, providing him with information about
the code base \emph{could} be a good thing even if not pertinent to the
current task.
However, \emph{none} of the side-topics actually led to \dFour
understanding something he did not already know (not shown above),
so we characterize this as a case of a
pair running into trouble (as defined in Section~\ref{sec:good-and-bad}).
Instead, the main topic (i.e., explaining the target state to \dFour) is
interrupted by twelve(!) abrupt topic changes
(only two of which are shown above).
Finishing an exchange with a net~time of 30 seconds takes the pair about
ten minutes---and it could have been even worse, as \dThree was nearly lost
after five minutes and only found his way back because a stacktrace happened to 
be displayed on-screen:
\begin{excerpt}
    \qa{\dThree}{OK. Now, where were we? [looks around, sees stacktrace
    on lower display corner] Ah, the exception, right.}
\end{excerpt}

\subsection{Anti-Pattern: Losing the Partner}
\label{sec:pat-losing}

Sometimes, one pair member is deeply engaged with the task at hand, trying to
understand the code or developing a design idea, but does not pay much attention
to her partner's
state of mind who then may or may not understand what the colleague is doing.
Such behavior may be \emph{expedient} in the short-term if, say, a defect is
found sooner than later, but it reduces the \emph{Togetherness} of the pair
which may result in
(a)~the partner being less knowledgable later or
(b)~missing a learning opportunity
(as discussed in \cite[Sec.~6.4.3]{ZiePre20-ppsessiondyn}).

\textbf{Example 2: Session \textsl{CA1} (19:00--21:00).}
Developer \cOne already started implementing a new form when \cTwo joins.
They want to make the form interactive such that one checkbox deactivates
multiple input fields (called ``panels'' and ``components'' below).
\cTwo appears to see problems with their approach but neither
explains them to \cOne nor reacts to \cOne's questions:
\begin{excerpt}
    \qa{\cTwo}{The problem is, it doesn’t fit with \emph{getComponents}
    [scrolls through file]}

    \qb{\cOne}{Why doesn’t it fit?}

    \qa{\cTwo}{I think so. I could be wrong. [continues scrolling]}

    \qb{\cOne}{We only need to get the individual component from the
    panel, right? Is that complicated?}

    \qa{\cTwo}{[ignoring C1] Ah, I just see it has a \emph{getContent}.}

    \qb{\cOne}{[reading from screen] A \emph{PanelBuilder}. Can we
    possibly get the other panels from there?}

    \qa{\cTwo}{[ignoring C1, continues scrolling] I’m not sure whether
    this all will work.}

    \qb{\cOne}{Can we deactivate a \emph{JPanel} on its own?}

    \qa{\cTwo}{[ignoring C1, continues scrolling] OK, I’d say---Shall we
    simply try to implement the methods?}

    \qb{\cOne}{Yeah, sure.}
\end{excerpt}

Although the pair appears to have reached an agreement
(\textcolor{devOneClr}{``Shall we?''}---\textcolor{devTwoClr}{``Yeah, sure.''}),
\cTwo has been \emph{Losing his Partner} during the above two minutes;
there is no way in which \cOne could have properly assessed
the proposal he agreed to, given that every question he asked was ignored.
Similar behavior of \cTwo occurs multiple times in session \textsl{CA1}.

\subsection{Anti-Pattern: Drowning the Partner}
\label{sec:pat-drowning}

The opposite behavior is also a problem.
Just as one pair member may provide too little explanation, she may also
\emph{Drown the Partner} in too many explanations that
(a)~go far beyond the task and are hence not \emph{expedient} and
(b)~also threaten the pair's \emph{Togetherness}.

\textbf{Example 3: Session \textsl{PA3} (29:50--31:40).}
Developers \pOne and \pThree just extracted multiple occurrences of the
value \texttt{0.01} that is used in several percentage calculations into a constant.
\pOne starts a long-winded explanation which his partner does not understand
because it is built on hypotheticals and does not relate to their actual code changes.
\pThree gets increasingly frustrated over the course of two minutes:
\begin{excerpt}
    \qa{\pOne}{It’s important to make clear that the last two `\emph{0.01}' have no relationship.}

    \qb{\pThree}{Which last two?}

    \qa{\pOne}{The last two in lines 31 and 32, for example.
    Assuming the two numbers would have no relation and someone who only sees the
    implementation with raw numbers thinks ‘Oh, there is a relation,
    I’ll introduce a constant’ [...].}

    [... \pOne goes on for 40 seconds ...]

    \qb{\pThree}{But applied to our case this has no relevance.}

    \qa{\pOne}{Yes, it has. Because it is a Magic Number, and Magic Number means}---%
    \qb{\pThree}{But it is no longer ‘magic’. We just named it!}

    \qa{\pOne}{I wanted to explain why we are doing this}---%
    \qb{\pThree}{[annoyed] I got that.}---%
    \qa{\pOne}{I only want to clarify that it’s important to}---%
    \qb{\pThree}{[annoyed, staring at screen] Got it.}---%
    \qa{\pOne}{make the relation with this renaming. [...] Not only to rename the variable.}
    \qb{\pThree}{[annoyed] It’s ok.}
\end{excerpt}
In general, these two developers are getting along great, but here
\pOne was \emph{Drowning his Partner} with
explanations the partner neither wanted
nor needed.
After the session, the two developers talked about that incident.
\pThree criticized \pOne for providing such
\textcolor{devTwoClr}{\emph{``unwanted lectures''}} too often,
continuing
\textcolor{devTwoClr}{\emph{``If I didn’t know you better, I’d perceive this behavior as arrogant''}}.
\pOne responded that some issues just need pro-active explanations, because the partner would not even know what to ask, or when.
Both agreed on this and continued working productively the next day.

\subsection{Doing the Right Thing}
\label{sec:positive}

Good pairs maintain \emph{Togetherness} and \emph{Expediency};
they avoid the three negative patterns described above.

For the case of \emph{Getting Lost in the Weeds}, both pair members are
responsible for restraining their impulses of following new ideas right away.
If good pairs get side-tracked, they notice it and work their way back together, as in the next example.

\textbf{Example 4: Session \textsl{JA1} (04:00--06:40).}
Early in the session, a simple question by \jOne interrupts an
explanation by \jTwo, which leads to a misunderstanding that takes almost
two minutes to clear up.
To avoid \emph{Getting Lost in the Weeds}, both partners explicitly switch back to the original topic (last two lines in the excerpt):
\begin{excerpt}
    \qa{\jTwo}{There is the central [News] plugin and multiple processors which each handle one wave. [...] It checks how the file size is changing.}

    \qb{\jOne}{In what time window are you looking?}

    [... two minutes of misunderstanding ...]

    \qb{\jOne}{\emph{30 seconds}, that’s what I wanted.}

    \qa{\jTwo}{That’s 30 seconds long, the time window. Now I got you.
    I can show it to you [in the code] in a minute.}

    \qb{\jOne}{Yes. And the NewsPlugin is doing what in all of this? Does it do exactly this monitoring and the delegation to the individual wave plugins, or what?}

    \qa{\jTwo}{No. The NewsPlugin basically only does---it gets called periodically by the cron server [...]}
\end{excerpt}

In contrast to \emph{Getting Lost in the Weeds}, which the partners do together,
\emph{Losing} or \emph{Drowning the Partner} is asymmetrical:
One pair member is doing something `wrong' to her partner
(i.e., explaining too little or too much) who should then try to avert the
problem.
The next two examples show how good pairs agree on which topics to address and
how to limit the scope of an explanation.

\textbf{Example 5: Session \textsl{DA2} (1:30:50--1:38:00).}
After their somewhat chaotic start (see Example~1
in Section~\ref{sec:pat-weeds}),
the session of \dThree and \dFour proceeds more orderly.
Newly hired \dFour explicitly asks his partner multiple
times whether he should provide an explanation on some matter, which \dThree
agrees to:
\begin{excerpt}
    \qb{\dFour}{Do you know about OSGi class loading?}

    \qa{\dThree}{Class-what? Not really, no.}

    \qb{\dFour}{Should I tell you?}

    \qa{\dThree}{Sure.}
\end{excerpt}

\textbf{Example 6: Session \textsl{JA1} (13:15--13:45).}
\jTwo explains code he wrote earlier to \jOne, who limits the explanation's scope:
\begin{excerpt}
    \qa{\jTwo}{It comes from the function \emph{`getLastFile()'}. [...] Shall
    we look into the function or not?}

    \qb{\jOne}{No, not now, please.}

    \qa{\jTwo}{Not now, ok.}
\end{excerpt}

In both cases, the developers make sure to neither \emph{Lose} nor \emph{Drown}
their partner.

\subsection{Further Elements of Pair Programming Skill}

We found more elements of pair programming skill which we do not report here,
e.g., \emph{Agreeing on a Common Plan}, which includes picking up on cues that
the partner did not understand one's ideas.
Failure to do so threatens the pair's \emph{Togetherness} and their work's
\emph{Expediency}.

\section{Discussion}
\label{sec:discussion}

Previous research and practitioner reports suggest that getting accustomed to
working in pairs and getting familiar with a particular partner takes time.
However, exhibited PP skill does not appear to directly depend on experience:
We saw developers with \emph{no PP experience} skillfully avoid
\emph{Losing} or \emph{Drowning} their partner,
e.g., \dFour probing his partner's knowledge gaps in the latter half of session \textsl{DA2} (which was \dFour's very first PP session ever), such as in Example~5;
and we saw problematic behavior in developers with long PP histories,
e.g., \cTwo (6 years of PP) \emph{Losing his Partner} in session \textsl{CA1},
or \pOne (2 years of PP) \emph{Drowning his Partner} in session \textsl{PA3},
see Examples~2 and 3.

This leads to open questions:
How do PP novices manage to have good PP sessions?
Which elements of PP skill can be acquired through what types of experience?
Which are specific to the context and the involved partners?

\section{Summary and Further Work}
\label{sec:conclusion-further-work}

Pair programming  (PP) does not just `work' because two software developers
sit next to each other.
Rather, developers can be more or less \emph{skilled} at pair programming.
We characterize two elements of that skill that are independent from software
development skills:
Maintaining \emph{Togetherness} and keeping an eye on \emph{Expediency}.
There are possibly more elements to be found.
So far, we described three patterns of problematic behavior:
\begin{itemize}
    \item \emph{Getting Lost in the Weeds}, during which both partners stay
    together as a pair, but lose sight of which topics are worth pursuing.

    \item \emph{Losing the Partner}, in which one pair member focuses too much
    on the task and explains too little.

    \item \emph{Drowning the Partner}, in which one pair member explains too
    much, which may harm the pair's Togetherness \emph{and} Expediency.
\end{itemize}

Our current data is limited to snapshots of pair programmer behavior:
Most pairs in the \emph{PP-ind} repository
\cite{ZiePre20-ppind-tr} were recorded only once or twice during a short stretch
of their pair programmer career.
Nevertheless, we have already seen that prior PP experience alone does not
explain beneficial and problematic behavior.
Longitudinal research with the same developers over longer time frames
will be needed to understand and disentangle the influence of
developers' personal styles,
their day-to-day form, and their experience with pair programming in
general or with a particular partner.

\section{Data Availability}

Unfortunately, we cannot share the full audio and video material due to 
non-disclosure agreements with the involved companies and confidentiality 
agreements with the recorded developers.
We do, however, provide the full original German transcripts as well as English
translations for the referenced session excerpts \cite{PPind-1.0}.

\printbibliography

\end{document}